\documentclass[prd,twocolumn,aps, floatfix,preprintnumbers,showpacs,nofootinbib,superscriptaddress]{revtex4-1}

\usepackage{graphicx} 
\usepackage{dcolumn} 
\usepackage{bm}        
\usepackage{amssymb}
\usepackage{color} 
\usepackage{comment}

\definecolor{darkgreen}{rgb}{0,0.7,0}

\newcommand\aj{\textit{AJ}}%                                         % Astronomical Journal
\newcommand\apjl{\textit{ApJ}}%                                      % Astrophysical Journal, Letters
\newcommand\apjs{\textit{ApJS}}%                                     % Astrophysical Journal, Supplement
\newcommand\aap{\textit{A$\&$A}}%                                    % Astronomy and Astrophysics
\newcommand\mnras{\textit{MNRAS}}%                                   % Monthly Notices of the RAS
\newcommand\physrep{\textit{Phys.~Rep.}}%                            % Physics Reports
\newcommand\jcap{\textit{JCAP}}%      
\newcommand\na{\textit{New Astronomy}}%      

\begin{document}

%%%%%%%%%%%%%%%%%%%%%%%%%%%%%%%%%%%%%%%%%%%%%%%%%%%%%%%%%%%%%%%%%%%%%%%%%%%%%%%%%%%%%%%%%%%%%%%%%%%%%%%%%%%%%%%%%%%%%%%%%%%%%%%%%%%%%%%%%%%%%%%%%%%%%%%%%%%%%%%%%

% INTERNAL INFO 

\widetext

%%%%%%%%%%%%%%%%%%%%%%%%%%%%%%%%%%%%%%%%%%%%%%%%%%%%%%%%%%%%%%%%%%%%%%%%%%%%%%%%%%%%%%%%%%%%%%%%%%%%%%%%%%%%%%%%%%%%%%%%%%%%%%%%%%%%%%%%%%%%%%%%%%%%%%%%%%%%%%%%%

% TITLE & AUTHORS

\title{Constraints on dark radiation from cosmological probes}

\author{Graziano Rossi}
\email{graziano@sejong.ac.kr}
\affiliation{Department of Astronomy and Space Science, Sejong University, Seoul, 143-747, Korea}

\author{Christophe Y{\`e}che}
\affiliation{CEA, Centre de Saclay, Irfu/SPP, F-91191 Gif-sur-Yvette, France}
  
\author{Nathalie Palanque-Delabrouille}
\affiliation{CEA, Centre de Saclay, Irfu/SPP, F-91191 Gif-sur-Yvette, France}

\author{Julien Lesgourgues}
\affiliation{CERN, Theory Division, CH-1211 Geneva 23, Switzerland}
\affiliation{LAPTh, Univ. de Savoie, CNRS, B.P.110, Annecy-le-Vieux F-74941, France}       

\date{\today}

%%%%%%%%%%%%%%%%%%%%%%%%%%%%%%%%%%%%%%%%%%%%%%%%%%%%%%%%%%%%%%%%%%%%%%%%%%%%%%%%%%%%%%%%%%%%%%%%%%%%%%%%%%%%%%%%%%%%%%%%%%%%%%%%%%%%%%%%%%%%%%%%%%%%%%%%%%%%%%%% 

% ABSTRACT

\begin{abstract}

We present joint constraints on the number of effective neutrino species $N_{\rm eff}$ and the sum of neutrino masses $\sum m_{\rm \nu}$,
based on a technique which exploits the full information contained in the 
one-dimensional Lyman-$\alpha$ forest flux power spectrum, complemented by additional cosmological probes.
In particular, we obtain $N_{\rm eff}= 2.91^{+ 0.21}_{- 0.22}$ (95\% CL) 
and $\sum m_{\rm \nu} <0.15$ eV (95\% CL) when we combine BOSS Lyman-$\alpha$ forest data with CMB (Planck+ACT+SPT+WMAP polarization) measurements,
and $N_{\rm eff}= 2.88 \pm 0.20$ (95\% CL) 
and $\sum m_{\rm \nu} <0.14$ eV (95\% CL) when we further add baryon acoustic oscillations. 
Our results provide strong evidence for the Cosmic Neutrino Background from $N_{\rm eff}\sim 3$ ($N_{\rm eff} = 0$ is rejected at more than $14~\sigma$), and 
rule out the possibility of a sterile neutrino thermalized with active neutrinos (i.e., $N_{\rm eff}$ = 4)  
-- or more generally any decoupled relativistic relic with $\Delta N_{\rm eff} \simeq 1$ -- 
at a significance of over  5 $\sigma$, the strongest bound to date, 
implying that there is no need for exotic neutrino physics in the concordance $\Lambda$CDM model.

\end{abstract}

%%%%%%%%%%%%%%%%%%%%%%%%%%%%%%%%%%%%%%%%%%%%%%%%%%%%%%%%%%%%%%%%%%%%%%%%%%%%%%%%%%%%%%%%%%%%%%%%%%%%%%%%%%%%%%%%%%%%%%%%%%%%%%%%%%%%%%%%%%%%%%%%%%%%%%%%%%%%%%%%%

% MAKETITLE

\pacs{CERN-PH-TH-2014-267}
\maketitle

%%%%%%%%%%%%%%%%%%%%%%%%%%%%%%%%%%%%%%%%%%%%%%%%%%%%%%%%%%%%%%%%%%%%%%%%%%%%%%%%%%%%%%%%%%%%%%%%%%%%%%%%%%%%%%%%%%%%%%%%%%%%%%%%%%%%%%%%%%%%%%%%%%%%%%%%%%%%%%%%%

% INTRODUCTION

\section{Introduction}

%---------------------------------------------------------------------------------------------------------------

The Standard Model of particle physics predicts that there are 
exactly three active neutrinos, one for each of the three charged leptons, and that neutrinos are all left-handed 
and with zero mass \cite{BER2012}. However, from experimental results on solar and atmospheric neutrino oscillations we now know that neutrinos are massive, with
at least two species being non-relativistic today \cite{MANG2005,LP2006,LMMP2013}. 
The distinctness of the three flavors, and the difference between neutrinos and antineutrinos depend critically on the  condition of being massless.
Therefore, the discovery that neutrinos have non-zero mass calls also into question the number of neutrino species \cite{CYB2004,MANG2007,MANG2011}.
All these issues have triggered an intense research activity in neutrino science over the last few years, 
with a remarkable interplay and synergy between cosmology and
particle physics. 
The measurement of the absolute neutrino mass scale remains the greatest challenge for both disciplines.
However, while particle physics experiments are capable of determining two of the squared mass differences, along with the number of active neutrino families, 
their mixing angles, and one of the complex phases \cite{CAP2014}, a combination
of cosmological  datasets allows one to place more competitive upper limits on the total neutrino mass (summed over the three families) as opposed to 
beta-decay experiments \cite{SEL2005,SEL2006,SIGNE2013,COS2014,PDB2015, PDB2015b,VGL2015}.
The knowledge of the total mass and type of hierarchy will complete the understanding of the neutrino sector, and shed light into several critical issues in
particle physics -- such as leptogenesis or baryogenesis. 

A variety of cosmological probes and complementary techniques can be used to study massive neutrinos, and to obtain 
stringent constraints on their total mass. 
The analysis of the cosmic microwave background (CMB) radiation provides the most direct route, 
especially via the early integrated Sachs-Wolfe (ISW) effect in polarization maps \cite{HIN2013,PLANCK2014cosmo,Planck_2015_parameters}
and  with gravitational lensing of the CMB by large-scale structure (LSS) \cite{SAN2013,BATT2014,BATT2015,PLANCK2014cosmo,Planck_2015_parameters}.
Other powerful LSS methods   include
the study of galaxy clusters with the  Sunyaev-Zel'dovich (SZ) effect, 
the determination of cosmic shear through weak lensing,  
the measurement of the three-dimensional matter power spectrum from galaxy surveys, and
 $21$ cm or Lyman-$\alpha$ (Ly$\alpha$) probes where the underlying tracer is neutral hydrogen (HI) \cite{KAIS1992,JAIN1997,ZALD1998,ABA2003}.
 In particular, remarkable progress has been recently achieved by exploiting the complementarity of the Ly$\alpha$ forest -- i.e. the absorption lines 
in the spectra of high-redshift quasars, due to HI in the intervening  photoionized inter-galactic medium (IGM) -- with other cosmological probes. 
This has been possible thanks to
extensive data provided by the Sloan Digital Sky Survey (SDSS)  \cite{SDSS2000}, which has dramatically increased the 
statistical power of the forest. 
Several studies in the literature have exploited the Ly$\alpha$ forest constraining power, mainly 
due to independent systematics and contrasting directions
of degeneracy in parameter space \cite{CROFT1998,CROFT2002,ZALD2001,SEL2005,SEL2006,MCDON2005,MCDON2006,KIM2008,VIEL2006,VIEL2010,PDB2013};
the synergy with distinct datasets has contributed to obtain competitive upper bounds of the total neutrino mass \cite{SEL2006,SIGNE2013,COS2014,PDB2015, PDB2015b,VGL2015}.

Cosmological measurements are also capable of constraining
the properties of relic neutrinos, and possibly of other
light relic particles  \cite{MANG2005,LP2006,MANG2011,LMMP2013}.
In particular, the 
density of radiation $\rho_{\rm R}$ in the Universe (which includes photons and additional species)
is usually parameterized by the effective number of neutrino species $N_{\rm eff}$,
and the neutrino contribution to the total radiation content is expressed in terms of $N_{\rm eff}$
via the relation
\begin{equation}
\rho_{\rm R} = \rho_{\rm \gamma}+\rho_{\rm \nu} = \Big [ 1 + {7 \over 8} \Big ( {4 \over 11} \Big )^{4/3} N_{\rm eff} \Big ] \rho_{\rm \gamma}, 
\label{eq1}
\end{equation}
where $\rho_{\rm \gamma}$ and $\rho_{\rm \nu}$ are the energy density of photons and neutrinos, respectively \cite{LP2006}.
This relation is valid when neutrino decoupling is complete, and holds as long as all neutrinos are relativistic.
In the Standard Model, $N_{\rm eff} =3.046$ due to non instantaneous
decoupling corrections (i.e., this corresponds to three active neutrinos, namely $N_{\rm \nu}=3$), and therefore any departure from this value would indicate non-standard
neutrino features or an extra contribution from other relativistic relics. 
Recently, there has been some mild preference for $N_{\rm eff} > 3.046$
from CMB anisotropy measurements \cite{KOM2011,HIN2013,HOU2014}: an excess from the expected standard number
could be produced by sterile neutrinos, a neutrino/anti-neutrino asymmetry or
any other light relics in the Universe; however, the latest results from Planck (2015) \cite{Planck_2015_parameters} combined with further
astrophysical data reported a value of $N_{\rm eff}$ consistent with that predicted by the Standard Model.  

To this end, Big Bang Nucleosynthesis (BBN) is a powerful tool for studying neutrino properties, 
as it accurately predicts the primordial light element abundances (i.e., deuterium, helium and lithium). 
Since the effective number of neutrino species 
parametrizes the expansion rate of the early Universe, precise measures of primordial abundances 
can provide stringent bounds on $N_{\rm eff}$ when combined with a measure of the baryon density
$\Omega_{\rm b}h^2$ obtained from the CMB \cite{CYB2015}. In turn, standard BBN (SBBN), which assumes  
microphysics characterized by Standard Model particle content and interactions with three light neutrino species, is a powerful probe 
for constraining physics beyond the Standard Model. 
In this respect, 
while in principle the $^4 {\rm He}$  mass fraction is a very sensitive probe of
additional light degrees of freedom as pointed out long ago \cite{STEIG1977}, systematic uncertainties severely limit its cosmological use; 
nevertheless, \cite{MANG2011} were able to derive $N_{\rm eff}$ constraints directly from BBN, bypassing the CMB, 
and reported a robust $N_{\rm eff}$ bound using $^4{\rm He}$ measurements alone. Their findings support the fact that extra radiation is strongly disfavored,
if not excluded, by BBN -- as we confirm in this work. Other results along these lines can be found in \cite{NOLL2015}.
Instead, as noted in   \cite{CYB2004}, 
the primordial deuterium ratio can provide more competitive bounds on $N_{\rm eff}$ when combined with CMB data. For instance, \cite{COO2014} reported a value $N_{\rm eff}=3.28 \pm 0.28$
from a combination of CMB observations and a novel measurement of primordial deuterium obtained from quasar absorption systems, provided that 
the values of $N_{\rm eff}$ and of the baryon-to-photon ratio did not change between BBN and recombination \cite{HOU2013}. Their result is consistent with 
Standard Model physics, and does not require additional sterile neutrinos, as we find in this study  with a different technique and robust statistical significance. 
The number of effective neutrino species can also be constrained with several other late-time LSS probes; see for example \cite{SANC2014}, 
where angle-averaged correlation functions and the clustering wedges measured from SDSS galaxies
are used to constrain $N_{\rm eff}$, in combination with CMB data, Type Ia supernovae, and Baryon Acoustic Oscillations (BAO) from other samples.
In addition, forecasts for Stage IV CMB polarization experiments about $N_{\rm eff}$ and other
cosmological parameters can be found in  \cite{WU2014}, and future prospects for the quantification of neutrino properties through the Cosmic Neutrino Background (CNB)
are reported in \cite{ABAZ2015}. 

In this paper, we present a new method to obtain joint constraints on $N_{\rm eff}$ and the total neutrino mass $\sum m_{\rm \nu}$
using the information contained in the 
one-dimensional Ly$\alpha$ forest flux power spectrum, complemented by other cosmological probes. 
The work carried out here further extends the technique used in \cite{PDB2015} along two directions: 
by considering joint constraints on $N_{\rm eff}$ and $\sum m_{\rm \nu}$ (an aspect not addressed in \cite{PDB2015}), and by expanding the 
global likelihood to accommodate non-standard dark radiation models via a novel analytic approximation (tested on a new set of non-standard
cosmological hydrodynamical simulations with $N_{\rm eff}$ different from its canonical value). 
In particular, we show how this technique is able to rule out
the presence of an additional sterile neutrino thermalized with three active neutrinos (i.e., $N_{\rm eff}=4$)  --
 or more generally any dark radiation -- at 
a significance of over  5 $\sigma$, and
provide strong evidence (greater than $14~\sigma$) for the CNB from $N_{\rm eff} \sim 3$.
Our results have important implications in cosmology and particle physics, especially suggesting that there is no indication for
 extra relativistic degrees of freedom,
and that the minimal $\Lambda$CDM model does not need to be extended further to accommodate non-standard dark radiation.

The paper is organized as follows. In Section \ref{sec:data}, we describe the various 
datasets adopted in this work with a particular emphasis on the Ly$\alpha$ forest sample. 
In Section \ref{sec:sims}, 
we  briefly illustrate our suite of hydrodynamical simulations with massive neutrinos 
used in the study, and explain how neutrinos are numerically implemented.
In Section \ref{sec:methods}, we outline our general technique to construct the global likelihood and 
obtain joint constraints on cosmological parameters; we also present our analytic approximation for the Ly$\alpha$ likelihood to
account for non-standard dark radiation scenarios, test the accuracy of the approximation in the linear regime, and explain in detail our analysis methodology -- i.e., frequentist versus Bayesian approach.
Joint constraints on $N_{\rm eff}$ and $\sum m_{\rm \nu}$ are presented in Section \ref{sec:results}, where we also test the validity of our approximation in the 
nonlinear regime via cosmological hydrodynamical
simulations with non-standard $N_{\rm eff}$ values.  
We conclude in Section \ref{sec:conclusions}, where we highlight the major achievements of this work, discuss their implications  
in cosmology and particle physics, and indicate future research directions.

%%%%%%%%%%%%%%%%%%%%%%%%%%%%%%%%%%%%%%%%%%%%%%%%%%%%%%%%%%%%%%%%%%%%%%%%%%%%%%%%%%%%%%%%%%%%%%%%%%%%%%%%%%%%%%%%%%%%%%%%%%%%%%%%%%%%%%%%%%%%%%%%%%%%%%%%%%%%%%%%%

% DATASETS

\section{Datasets}  \label{sec:data}

%---------------------------------------------------------------------------------------------------------------

The joint constraints on $N_{\rm eff}$ and $\sum m_{\rm \nu}$ presented in this work are obtained from a combination of LSS and CMB measurements.

As LSS probes, we used the one-dimensional Ly$\alpha$ forest flux power spectrum derived from the Data Release 9 (DR9) of the 
Baryon Acoustic Spectroscopic Survey (BOSS \cite{SDSS2000,BOSS2013}) quasar data \cite{PDB2013}, 
combined with the measurement of the BAO scale in the clustering of galaxies from  the BOSS Data Release 11 (DR11)  \cite{AND2014}.
BOSS \cite{BOSS2013} is the cosmological counterpart of the third generation of the SDSS, the leading ground-based astronomical survey designed to explore the large-scale distribution of galaxies and quasars by using a dedicated $2.5$m 
telescope at Apache Point Observatory \cite{SDSS2000}.
Specifically for the Ly$\alpha$ forest, our data consist of 13 821 quasar spectra, carefully selected according to their high quality, signal-to-noise ratio and spectral resolution, to bring systematic uncertainties at the same level of the statistical uncertainties. 
The Ly$\alpha$ forest flux power spectrum is measured in twelve redshifts bins, from $\langle z \rangle =2.2$ to $4.4$, in intervals of $\Delta z =0.2$, and
spans thirty-five wave numbers in the $k$ range  $[0.001 - 0.02]$, with $k$ expressed in ${\rm (km/s)}^{-1}$, which corresponds approximately to $[0.1 - 2]$ (Mpc/h)$^{-1}$ at $z \sim 3$.
Correlations between different redshift bins were neglected, and the Ly$\alpha$ forest region was divided into up to three distinct $z$-sectors to minimize their impact. 
Noise, spectrograph resolution, metal contaminations and other systematic
uncertainties were carefully subtracted out or accounted for in the modeling \cite{PDB2013}. 

As CMB probes, we adopted a combination of datasets collectively termed `CMB', which includes
Planck (2013) temperature data from the March 2013 public release (both high-$\ell$ and low-$\ell$) \cite{PLANCK2014likelihood}, 
the high-$\ell$ public likelihoods from the Atacama Cosmology Telescope (ACT) \cite{ACT2014} and the South Pole Telescope (SPT) \cite{SPT2012} experiments, and
some low-$\ell$  WMAP polarization data  \cite{WMAP2013}.
 
%%%%%%%%%%%%%%%%%%%%%%%%%%%%%%%%%%%%%%%%%%%%%%%%%%%%%%%%%%%%%%%%%%%%%%%%%%%%%%%%%%%%%%%%%%%%%%%%%%%%%%%%%%%%%%%%%%%%%%%%%%%%%%%%%%%%%%%%%%%%%%%%%%%%%%%%%%%%%%%%%

% SIMULATIONS

\section{Simulations}  \label{sec:sims}

To constrain neutrino masses and possible extra-relativistic degrees of freedom exploiting  Ly$\alpha$ forest information, 
a detailed modeling of the line-of-sight power spectrum of the Ly$\alpha$ transmitted flux  
is required. This is because the scales probed lie fully in the non-linear regime, and therefore
non-linear simulations are  necessary to compare to the Ly$\alpha$ one-dimensional flux power spectra 
(while the BAO peak scale is not relevant here). 
To this end, we 
devised a novel suite of hydrodynamical cosmological simulations which include massive neutrinos \cite{ROS2014} to map the parameter space
around the central reference model on a regularly-spaced grid, and   compute  
first and second-order derivatives in the Taylor expansion of the Ly$\alpha$ forest flux.
We use those simulations here in combination with the previously described datasets 
to obtain bounds on $N_{\rm eff}$ and $\sum m_{\rm \nu}$, together with an analytic approximation to include non-standard dark radiation scenarios in the Ly$\alpha$ likelihood. 
We also run several non-standard dark radiation simulations to test the validity of such approximation in the nonlinear regime.
In what follows, we first briefly review the basic characteristics of our simulation suite, and then 
digress on the numerical implementation of massive neutrinos. 

 %---------------------------------------------------------------------------------------------------------------

\subsection{Simulation Suite with Massive Neutrinos}
 
 We modeled  the nonlinear evolution of the gas, dark matter, and neutrinos with a 
smoothed particle hydrodynamics (SPH) Lagrangian technique in its `entropy formulation' \cite{GM1977,L1977,SH2002}, where
all the components are  treated individually as a set of separate particles \cite{SH2002}. 
The gas, photo-ionised and heated by a spatially uniform 
ionising background, is assumed to be of primordial composition with a helium mass fraction of $Y=0.24$, while
metals and evolution of elementary abundances are neglected. 
This
background was applied in the optically thin limit and switched on at $z=9$; the thermal history in the simulations is
consistent with the temperature measurements of \cite{BBHS2011} through an adaptation
of the cooling routines. We also explored a variety of different thermal histories, by rescaling the amplitude and
density dependence of the photoionization heating rates in the simulations. 
We used the same simplified criterion for star formation as in \cite{VIEL2010}, but 
improved on previous studies in several direction, in particular with 
updated routines for IGM radiative cooling and heating processes, and
initial conditions based on second-order Lagrangian perturbation theory (2LPT)
 rather than the {Zel'dovich} approximation.
We adopted Gadget-3 \cite{SPRING2001,SPRING2005}
 for evolving Euler hydrodynamical equations, primordial
chemistry with cooling and some externally specified ultraviolet (UV) background, 
supplemented by CAMB \cite{CAMB2000} and a modified version of 2LPT \cite{CROCCE2006} for determining 
the initial conditions. We disabled feedback options, and neglected galactic winds. 

For a given neutrino mass and various combinations of cosmological parameters,  we performed a set of three simulations with different box sizes and number of particles
appropriate for the quality of BOSS -- but readily 
adaptable  for upcoming or future experiments, such as eBOSS and DESI  \cite{DESI2013,FR2014,ABA2015}.
We assumed periodic boundary conditions, adopted a box size of $100~h^{-1}{\rm Mpc}$ for large-scale power  with a number of particles per component $N_{\rm p}=768^3$ 
and a box size of $25~h^{-1}{\rm Mpc}$ for small-scale power, in the latter case with $N_{\rm p}=768^3$ or $192^3$, respectively. 
Aside from the central cosmological simulation indicated as the `best guess' run, which has only a massless neutrino component, all our other
simulations contain three degenerate species of massive neutrinos with $M_{\rm \nu}=0.1,0.2,0.3,0.4,0.8$ eV, respectively. 
A splicing technique introduced by \cite{MCD2003}   is further used to 
achieve an equivalent resolution of 
 $3 \times 3072^3 \simeq 87$ billion particles in a $(100~h^{-1} {\rm Mpc} )^3$ box size 
-- reducing the resolution and thus the computational requirements of our numerical simulations. 
To this end, the small-scale neutrino clustering has also been neglected.

We started our runs at  $z = 30$, with 2LPT initial conditions  having the same random seed, and 
produced snapshots at regular intervals in redshift  between $z=4.6 - 2.2$, with $\Delta z=0.2$.
For each individual simulation, 100 000 skewers were drawn with random origin and direction, and the one-dimensional power spectrum computed at different redshifts.
The final theoretical power spectrum is an average obtained from all the individual skewers, for any given model. 
We acknowledge that, even though we used 2LPT initial conditions, an earlier starting redshift would have been more ideal  
because the neutrino background energy density is slightly 
relativistic and at early times deviates from $\Omega_\nu(a=1)/a^3$ -- with $a$ the expansion factor and $\Omega_{\rm \nu}$ the neutrino density -- thus altering the growth rate; our final choice for the starting redshift
was mainly a compromise between
the available computational time and the large number of simulations we had to perform for our grid-based technique.
For more technical details on the simulations, pipeline, and neutrino inclusion we refer the interested reader to \cite{ROS2014}.
 
 %---------------------------------------------------------------------------------------------------------------

\subsection{Numerical Implementation of Massive Neutrinos}
 
We modeled massive neutrinos as a separate set of particles in our simulations, similarly to what is routinely done
for the gas and the dark matter components when an SPH formulation is adopted \cite{GM1977,L1977,SH2002}.  
 A full hydrodynamical treatment is then carried out, well-inside the nonlinear regime, including the
effects of baryonic physics which affect the IGM -- resulting in a computationally intensive approach. 
Within the range of degenerate neutrino masses, 
their thermal velocities can be approximated as \cite{LP2006}
\begin{equation}
v_{\rm th} \sim 150(1+z) \Big[ {1 {\rm eV} \over \sum m_{\rm \nu} } \Big] ~{\rm km/s}.
\label{eq:nuvel}
\end{equation}
Clearly, given their high thermal velocities, modeling massive neutrinos numerically is
a nontrivial task, particularly because of significant shot-noise. 
However, resolving nonlinear scales is important for our detailed modeling of the small-scale flux power spectrum; to this end, 
using some accurate approximate linear solutions such as those proposed by  \cite{BH2010,AHB2013}
would help in speeding-up the calculations considerably, but we instead opted for a fully nonlinear $N$-body treatment in this work.  

The central element of our joint constraints on $N_{\rm eff}$ and $\sum m_{\rm \nu}$
is our simulation-based Taylor expansion model for the dependence of the
Ly$\alpha$ power spectrum on cosmological and astrophysical parameters.
In particular, as shown in \cite{PDB2015}, the small but significant scale dependence of the total 
matter power spectrum response is a consequence of nonlinear evolution that can only be
modeled accurately using hydrodynamical simulations having a neutrino component implemented as a separate set of particles:
this allows one to quantify the response of the power spectrum to isolated variations in individual parameters, 
and in particular  to disentangle the well-known degeneracy between $\sum m_{\rm \nu}$
and the power spectrum amplitude $\sigma_8$ \cite{LP2006,VIEL2010,LMMP2013}. 
However, what is really driving the neutrino mass constraints is the 
amplitude of the Ly$\alpha$ flux power spectrum at small scales, while the 
dependence of the flux power spectrum on $M_\nu$ with $\sigma_8$ fixed is less than 1\% for a 2$\sigma$ change (see Fig. 12 in \cite{PDB2015}),
compatible with the uncertainty associated with our numerical simulations and therefore not significant. 

Several alternative attempts to model neutrinos in numerical simulations,
either by using linear approximations, hybrid techniques,  or treating neutrinos as a fluid with a grid method can be found in 
\cite{WFD1983,BH2008,BH2009,BH2010,AHB2013} -- but those implementations are not further considered here.
  
%%%%%%%%%%%%%%%%%%%%%%%%%%%%%%%%%%%%%%%%%%%%%%%%%%%%%%%%%%%%%%%%%%%%%%%%%%%%%%%%%%%%%%%%%%%%%%%%%%%%%%%%%%%%%%%%%%%%%%%%%%%%%%%%%%%%%%%%%%%%%%%%%%%%%%%%%%%%%%%%%

% METHODOLOGY

\section{Methodology}  \label{sec:methods}      

%---------------------------------------------------------------------------------------------------------------

To derive joint constraints on $N_{\rm eff}$ and $\sum m_{\rm \nu}$, we
extended the procedure applied in \cite{PDB2015} by using an analytic approximation to include non-standard dark radiation models in the Ly$\alpha$ likelihood.
In what follows, we first describe our general technique to construct the global likelihood; we then clarify our frequentist-based analysis method, 
digress on the comparison between frequentist and Bayesian techniques, and finally elaborate on the
analytic approximation adopted when $N_{\rm eff}$ is different from the canonical expectation.

%---------------------------------------------------------------------------------------------------------------

\subsection{Multidimensional Likelihood Construction and Frequentist Analysis} \label{sec:frequentist}

Our main goal is to construct a multidimensional likelihood $\mathcal{L}$, which is the product of individual likelihoods 
defining the various cosmological probes considered (LSS and CMB), i.e., 
$\mathcal{L} =   \mathcal{L}^{\rm LSS}   \mathcal{L}^{\rm CMB} =
 \mathcal{L}^{\rm Ly \alpha}  \mathcal{L}^{\rm BAO} \mathcal{L}^{\rm Planck} \mathcal{L}^{\rm ACT} \mathcal{L}^{\rm SPT}  \mathcal{L}^{\rm WMAP}$. 
 Along the lines of \cite{PDB2015}, for the CMB likelihood we
 assumed the best-fit and covariance matrix directly from the Planck results \cite{PLANCK2014likelihood,PLANCK2014cosmo} in the case of 
a $\Lambda$CDM model extended to massive neutrinos and an arbitrary number of massless extra degrees of freedom,
 while 
we used the correlation matrix with a posterior based on BAOs from the official Planck (2013) chains
to account for $\mathcal{L}^{\rm BAO}$;  
 therefore, all the correlations between parameters 
are taken into account with our technique. 
We then constructed the Ly$\alpha$ forest likelihood with an elaborated
procedure briefly described as follows -- but see \cite{ROS2014,BOR2014,PDB2015} for all the numerical and data-oriented aspects.
In detail, for a model $\cal{M}$ defined by three categories of parameters -- cosmological ($\boldsymbol{\alpha}$), astrophysical ($\boldsymbol{\beta}$), nuisance  ($\boldsymbol{\gamma}$) -- globally indicated
with the multidimensional vector  $\boldsymbol{\Theta} = (\boldsymbol{\alpha}, \boldsymbol{\beta}, \boldsymbol{\gamma})$, and for a $N_{\rm k} \times N_{\rm z}$
dataset $\boldsymbol{X}$ of power spectra $P(k_{\rm i}, z_{\rm j})$
measured in  $N_{\rm k}$ bins in $k$ and $N_{\rm z}$ bins in redshift with experimental Gaussian errors $\sigma_{\rm i,j}$, with $\boldsymbol{\sigma}= \{ \sigma_{\rm i,j} \}$,  $i=1, N_{\rm k}$ and $j=1, N_{\rm z}$, the Ly$\alpha$ likelihood is written as:
\begin{equation}
\mathcal{L}^{Ly\alpha} (\boldsymbol{X}, \boldsymbol{\sigma}|\boldsymbol{\Theta}) = {\exp[- (\Delta^{\rm T} C^{-1} \Delta)/2]  \over  (2 \pi)^{{N_{\rm k} N_{\rm z} \over 2}} \sqrt{|C|} } ~\mathcal{L}_{\rm prior}^{Ly\alpha}(\boldsymbol{\gamma})  
\label{eq2}
\end{equation}
where $\Delta$ is a $N_{\rm k} \times N_{\rm z}$ matrix with elements $\Delta(k_{\rm i}, z_{\rm j}) = P(k_{\rm i}, z_{\rm j}) - P^{\rm th}(k_{\rm i}, z_{\rm j})$,  
$P^{\rm th}(k_{\rm i}, z_{\rm j}) $ 
is the predicted theoretical value of the power spectrum for the bin $k_{\rm i}$
 and redshift $z_{\rm j}$ given the parameters ($\boldsymbol{\alpha}, \boldsymbol{\beta}$) and computed from simulations \cite{ROS2014},  
$C$ is the sum of the data and simulation covariance matrices, and $\mathcal{L}_{\rm prior}^{Ly\alpha} (\boldsymbol{\gamma})$
accounts for the nuisance parameters, a subset of the parameters $\boldsymbol{\Theta}$.
For the baseline model, we considered five cosmological parameters $\boldsymbol{\alpha}$ in the context of the $\Lambda$CDM paradigm assuming flatness,  i.e. $\boldsymbol{\alpha}=$($n_{\rm s}, \sigma_8, \Omega_{\rm m}, H_0, \sum m_{\rm \nu}$),
four astrophysical parameters $\boldsymbol{\beta}$ related to the state of the  IGM -- two for the effective optical depth of the gas assuming         
a power law evolution, and two related to the heating rate of the IGM -- and 12 nuisance parameters $\boldsymbol{\gamma}$ to account for
imperfections in the measurements and in the modeling, plus
two additional parameters for the correlated absorption of Ly$\alpha$ and either Si-III or Si-II. 
The theoretical Ly$\alpha$ power spectrum $P^{\rm th}(k_{\rm i}, z_{\rm j})$, as a function of $\boldsymbol{\alpha}$ and $\boldsymbol{\beta}$,
is obtained via a second-order Taylor expansion around
a central model chosen to be in agreement with Planck (2013) cosmological results, and computed using the grid of simulations \cite{ROS2014}
previously described. 

The global likelihood $\cal{L}$  
is finally interpreted in the context of the frequentist approach \cite{NEY1937}.
This is done by minimizing the quantity $\chi^2(\boldsymbol{X}, \boldsymbol{\sigma}|\boldsymbol{\Theta}) = -2 \ln [\mathcal{L}(\boldsymbol{X}, \boldsymbol{\sigma}|\boldsymbol{\Theta})]$
for data measurements $\boldsymbol{X}$ with experimental Gaussian errors $\boldsymbol{\sigma}$. In particular, first we compute the global minimum
$\chi_0^2$, leaving all the $N$ cosmological parameters free; we then set confidence levels (CL)  on a chosen parameter
$\alpha_{\rm i}$ by performing the minimization for a series of fixed values of $\alpha_{\rm i}$ -- thus with $N-1$ degrees of freedom.
The difference between $\chi_0^2$ and the new minimum allows us to compute the CL on $\alpha_{\rm i}$. 
This technique is readily extended to higher dimensions, in order to derive joint constraints on two (or more) cosmological parameters.

When computing the uncertainties on the parameters, an exact profiling is done (frequentist equivalent to the Bayesian marginalization), by refitting all the parameters on every single point of the $\chi^2$ map:
we state uncertainty ranges obtained by letting all other parameters vary. 
Specifically, this means that nuisance parameters are allowed to take different values. The confidence intervals are thus directly comparable to other papers, as was demonstrated in
\cite{PDB2015} by a proper comparison between the frequentist approach used here and a full Bayesian approach. Such a comparison has also been studied by the Planck collaboration in \cite{Planck_2014_intermediate}. 
In particular, the uncertainty on $N_{\rm eff}$ is derived by letting $\sum m_{\rm \nu}$ vary as all other parameters (including nuisance parameters) -- equivalently to marginalizing over $\sum m_{\rm \nu}$. 
In addition, we require that $\sum m_{\rm \nu}>0$. Since the minimum of the fit occurs for a negative value of $\sum m_{\rm \nu}>0$, the $\chi^2$ difference is therefore computed with respect to the $\chi^2$ value for $\sum m_{\rm \nu}=0$. 
This approach is very similar to the prescription of Feldman-Cousins \cite{FC1998}, as verified in \cite{PDB2015}. 
We also note that, with respect to the CMB likelihood, the marginalisation is done using the full correlation matrix publicly released by the Planck collaboration, 
and asymmetric errors are also accounted for by the use of an asymmetric Gaussian width on either side of the maximum -- directly derived from the asymmetric 
uncertainties in the Planck full likelihood. This approach has been thoroughly tested in \cite{PDB2015}, and shown to give identical results, both in terms of central fit values, 
uncertainties and correlations between parameters, of a canonical Bayesian methodology based on the full Planck likelihood. 
We have thus chosen here the approach that is less time-consuming and more flexible, and adopted the same strategy in \cite{PDB2015b}. Next, we provide a concise comparison
between our selected statistical method and a more standard Bayesian interpretation. 

%---------------------------------------------------------------------------------------------------------------

\subsection{Frequentist versus Bayesian Interpretation}

A dispute between the frequentist approach (most common in particle physics) as opposed to Bayesian techniques (often adopted in cosmology) 
is beyond the scope of this paper. However, we would like to reiterate a few key concepts in support of our chosen interpretation methodology previously detailed.

The frequentist (or classical) method, originally introduced by \cite{NEY1937}, 
is based on a concept of probability that concerns the number of expected outcomes in a series of repeated tests, and is 
primarily focused on the probability of the data $\boldsymbol{X}$  given the parameters $\boldsymbol{\Theta}$, i.e., the likelihood $\mathcal{L} = p(\boldsymbol{X}|\boldsymbol{\Theta})$. 
In the frequentist sense, the best estimator of a parameter is the value of the parameter which maximizes the likelihood. 
Along with the global maximum likelihood, often the maximum likelihood at every fixed value of the parameter of interest is also given (i.e., the one-dimensional profile likelihood);
a maximum likelihood estimator which maximize  $\mathcal{L}$ is then identified. By definition, the frequentist approach does not require any marginalization to determine the 
sensitivity on a single parameter, and in principle no prior information is required. With this approach, correlations between variables are naturally encoded,
and the minimization fit can explore the entire phase space of parameters considered.

Bayesian techniques are based instead on a different concept of probability, intended as the `degree of belief' about a particular assertion.
Within this framework, the primarily focus is the posterior probability $p(\boldsymbol{\Theta}|\boldsymbol{X})$, namely the probability of the
parameters $\boldsymbol{\Theta}$ given the data $\boldsymbol{X}$.
The full posterior can be marginalized over some of the model parameters, to provide a posterior on the remaining parameters; this allows one to compute credible intervals for each single parameters, 
along with joint confidence contours on parameter pairs.  
 
The frequentist quantity of interest $p(\boldsymbol{X}|\boldsymbol{\Theta})$ and the Bayesian posterior $p(\boldsymbol{\Theta}|\boldsymbol{X})$
are related through the Bayes equation:
\begin{equation} 
p(\boldsymbol{X}|\boldsymbol{\Theta}) p(\boldsymbol{\Theta})  = p(\boldsymbol{\Theta}|\boldsymbol{X}) p(\boldsymbol{X}).
\end{equation}
 For problems of parameter inference, the normalizing Bayesian evidence may be neglected, and for the case where priors
are flat and fully enclose the likelihood the posterior is proportional to the likelihood, namely:
\begin{equation} 
p(\boldsymbol{X}|\boldsymbol{\Theta})  \propto p(\boldsymbol{\Theta}|\boldsymbol{X}).
\end{equation}
 In the latter case, both the posterior and the likelihood peak in exactly the same place in parameter space, and therefore both methods should provide an identical answer. 
 The slight controversy from the frequentist point of view lies in the necessity of specifying the 
prior probability of the unknown parameters $p(\boldsymbol{\Theta})$,  as it implies that some informed guess must be made
in advance of the collection of the observed data, concerning the plausible values of the unknown parameters. In this respect, 
priors must be carefully specified, and can be a delicate issue. On the other hand, the advantage
of being able to specify priors allows one to quantify skepticism about the quality of the experiment, or to test theoretical ideas or additional issues.  

In \cite{PDB2015} we have performed both a frequentist and a Bayesian analysis, 
and proved for a large number of configurations that we obtain
the same results using the two radically different approaches -- confirming the robustness of our parameter constraints. 
In particular, Table 4 in \cite{PDB2015} reports the results of a Bayesian analysis performed with 10 redshift bins and a free neutrino mass, when considering the Ly$\alpha$ forest alone;
the first column of the same table shows results when flat priors are assumed for all parameters (except for $H_0$, which has a Gaussian prior), and 
can be directly compared with the frequentist results in the last column of Table 3 in \cite{PDB2015}, obtained under the same assumptions. 
Similarly, for the combination Ly$\alpha$+CMB, the frequentist and Bayesian results can be compared through the fourth column of Table 5 in \cite{PDB2015} and the third column in Table 6 of the same paper.
Remarkably, the constraints obtained with the frequentist and the Bayesian techniques are very similar on all fit parameters,
with  only very minor differences in the final confidence limits. 
Hence, the  two approaches are in excellent agreement for central and $1 \sigma$ values, and therefore 
our bounds on cosmological parameters are robust even against a change of statistics;
the exact 2D contours may differ slightly but the accuracy of either approach is sufficient for
the scope of this study. 
As discussed also in  \cite{PDB2015},
the conceptual difference between the two methods should not   lead 
to major discrepancies in the estimate of physical parameters and their confidence intervals 
when the model parameters can be contained by the data, and confidence intervals obtained with both methods are thus directly comparable to other literature results. 
Due to the large number of nuisance parameters associated to the Planck, ACT/SPT and Ly$\alpha$ likelihoods, and to our extended parameter space,
we chose here the less time-consuming and more flexible approach, and adopted the same strategy in \cite{PDB2015b}. 
Before moving on to our main results, we briefly describe our analytic approximation 
used to incorporate non-standard radiation models in  $\mathcal{L}^{\rm Ly \alpha}$. 

%---------------------------------------------------------------------------------------------------------------

\begin{figure*}
\includegraphics[scale=0.885]{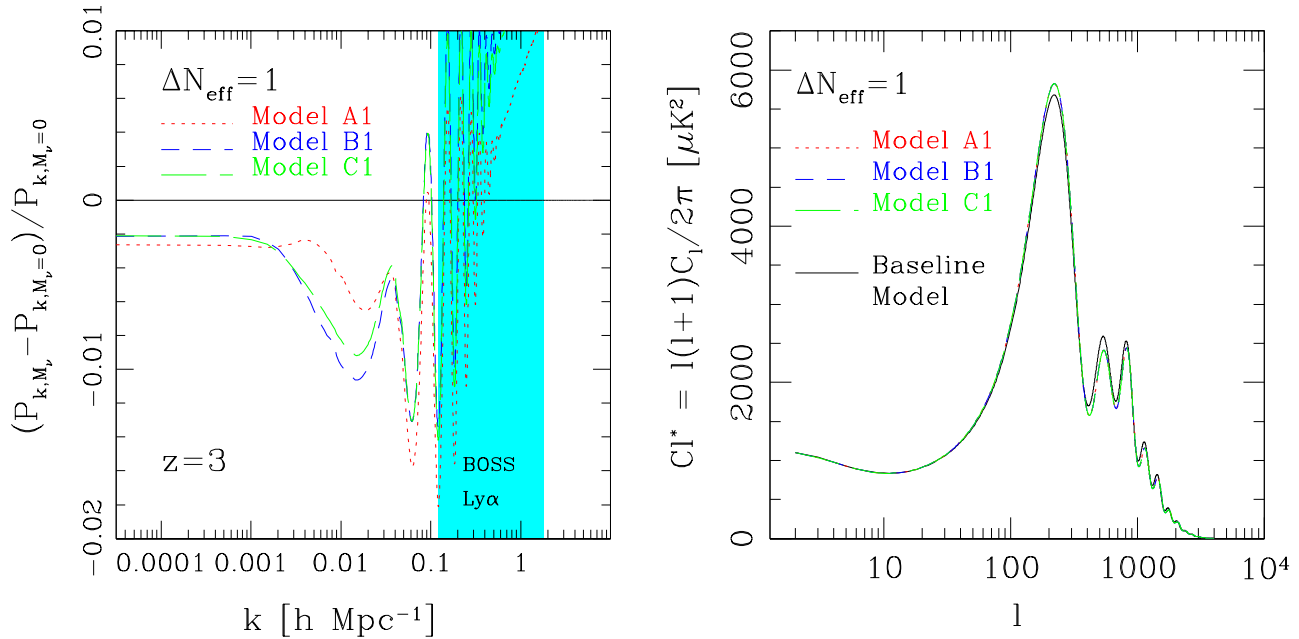}
\caption{Linear theory test of  the accuracy of our analytic approximation to include non-standard dark radiation models. [Left]
 Linear matter power spectra for a series of models $\tilde{\mathcal{M}}$ having $\Delta N_{\rm eff}=1$ at $z=3$ (chosen as a representative central value for the redshift range considered in this study), normalized by the
baseline model $\mathcal{M}$ with $N_{\rm eff}=3.046$ and three active neutrinos of degenerate mass, when $M_{\rm \nu}=0.3$ eV. See the main text for more details. 
[Right] Corresponding CMB temperature power spectra for the same models. Both panels show small differences in the scale of BAO and CMB peaks, but those differences do not affect the 
 Ly$\alpha$ likelihood.}
\label{fig1} 
\end{figure*}

%---------------------------------------------------------------------------------------------------------------

\begin{figure*}
\includegraphics[scale=0.38]{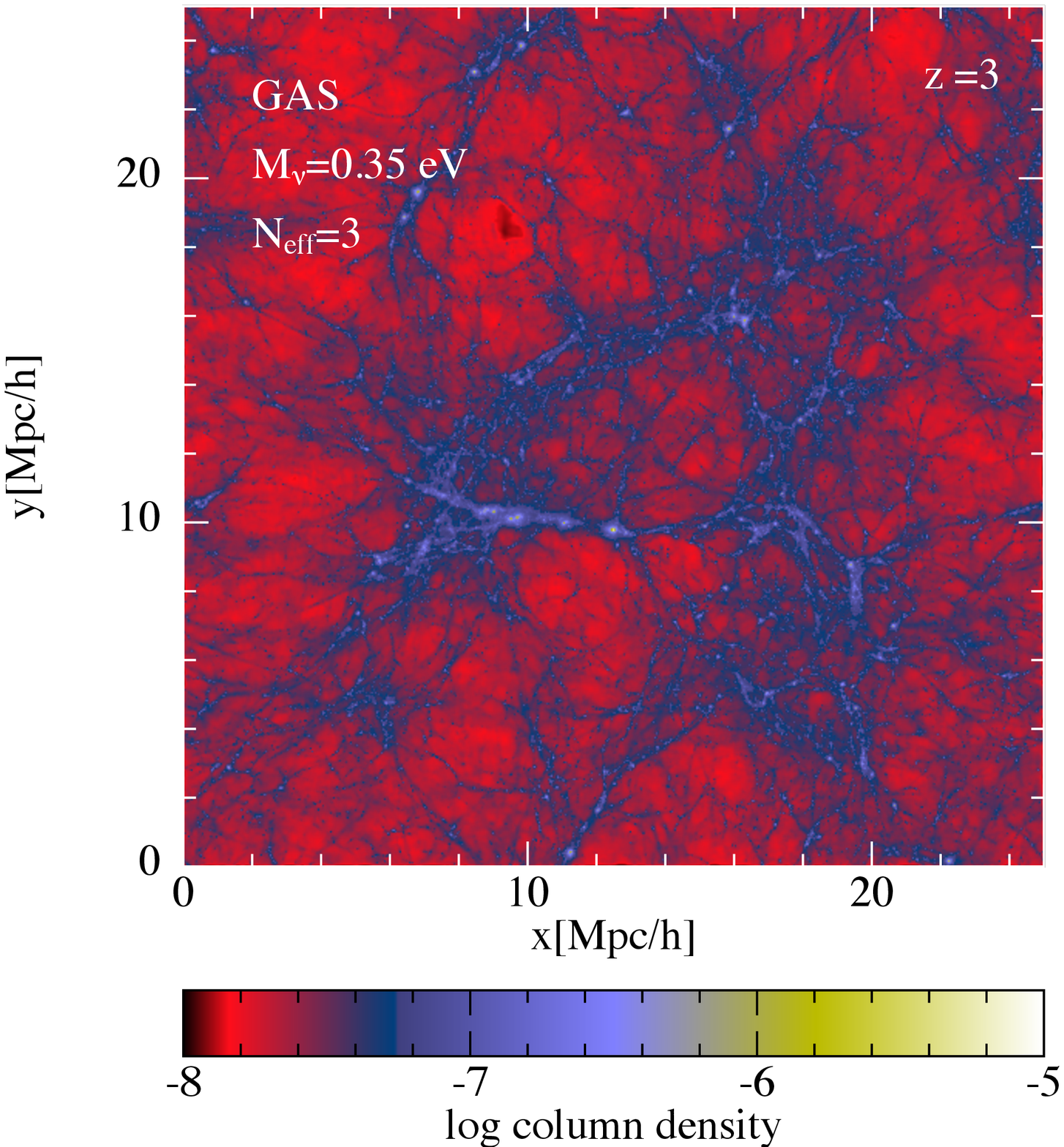}
\includegraphics[scale=0.38]{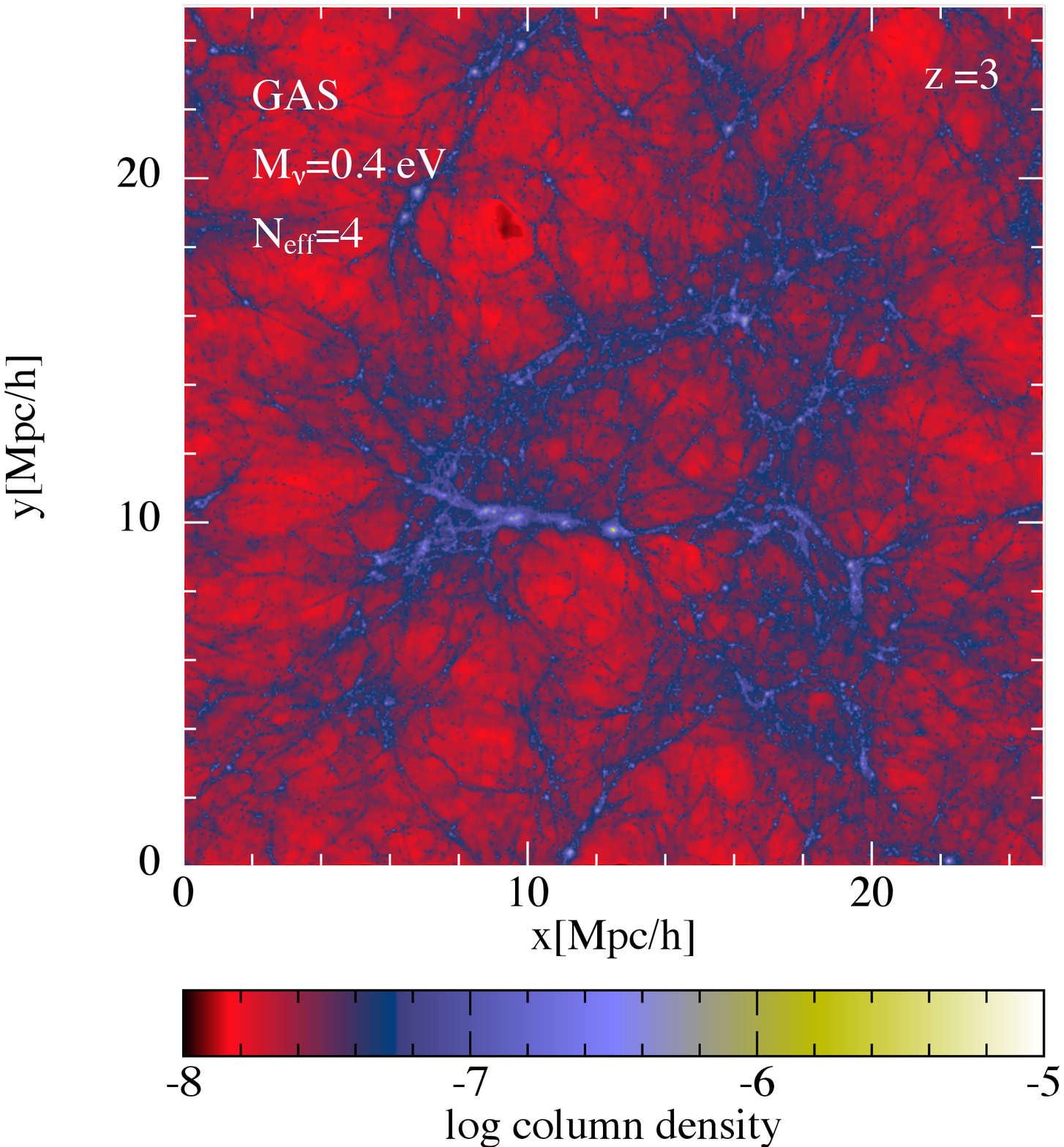}\\
\includegraphics[scale=0.38]{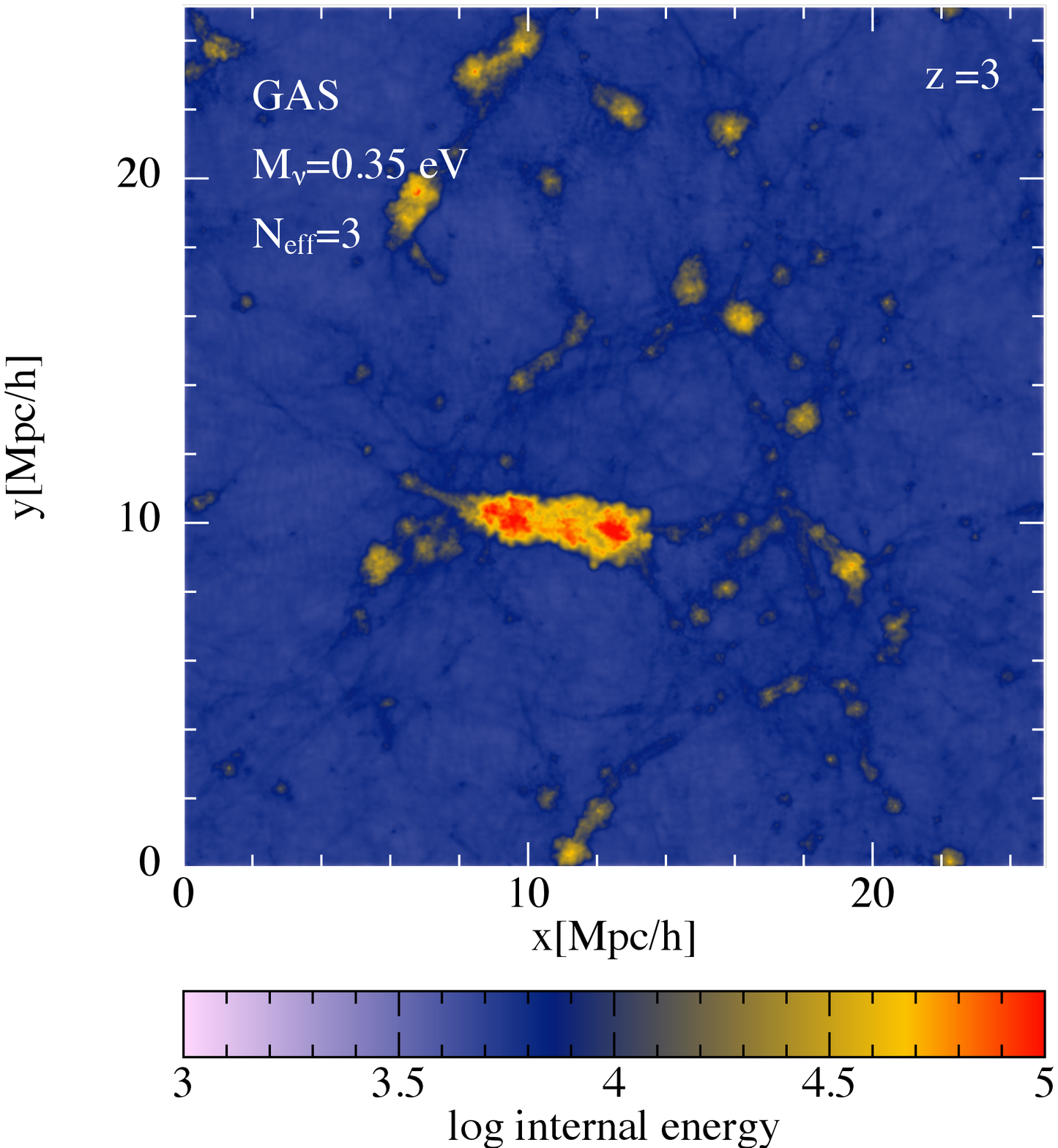}
\includegraphics[scale=0.38]{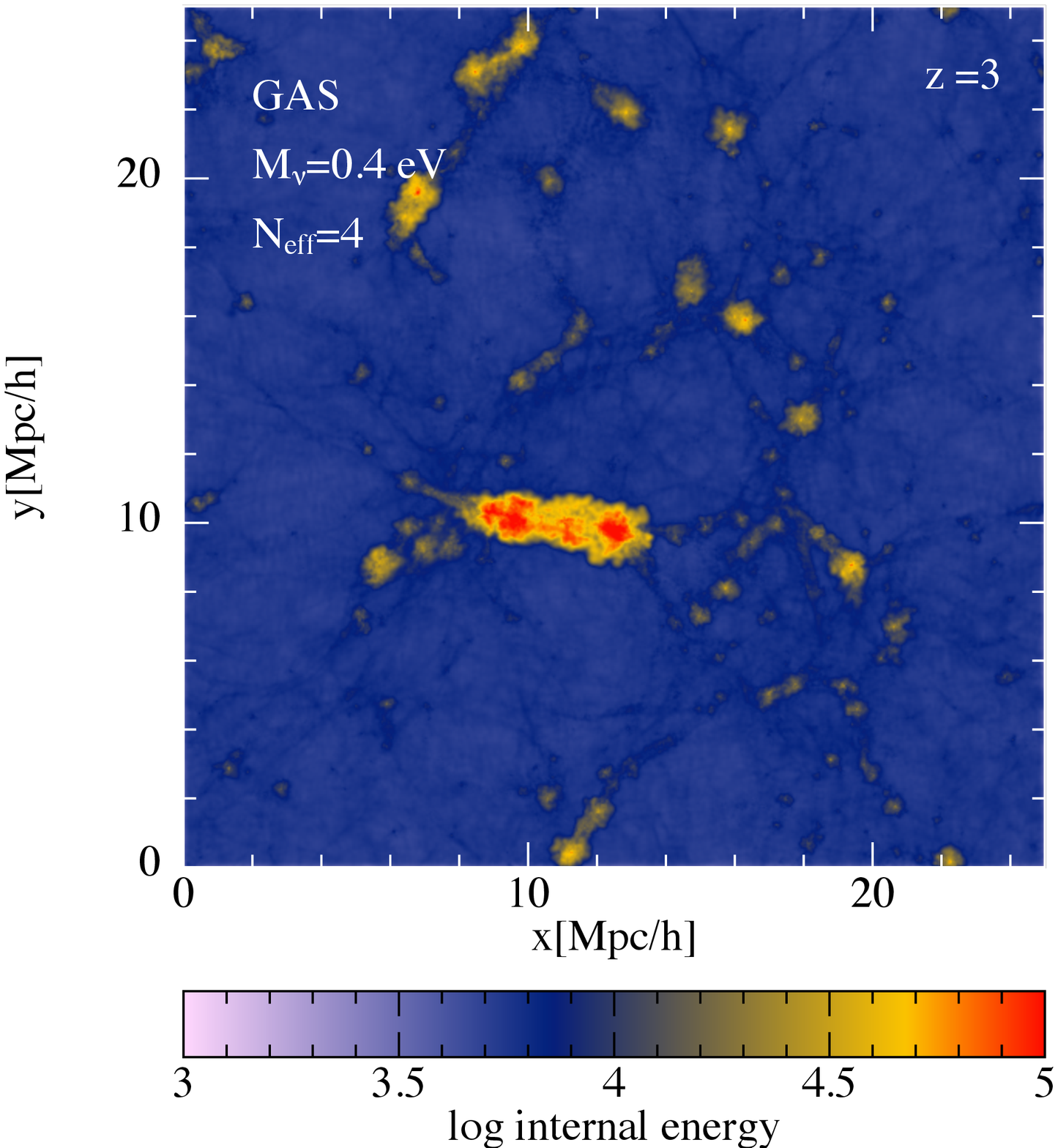}
\caption{Snapshots at $z=3$ from simulations with a canonical value of $N_{\rm eff}$ (left panels),
and when $N_{\rm eff}=4$ (right panels). The two cosmologies are related by our
analytic remapping: $\tilde{M}_{\rm \nu}=0.35$ eV for the baseline model, while  
$\tilde{M}_{\rm \nu}=0.4$ eV for the non-standard model 
which contains an additional massless sterile neutrino
assumed to be in thermal equilibrium with three degenerate active massive neutrinos. 
Top panels show projections of the gas density
along the $x$ and $y$ directions (and across $z$) for a
$25~h^{-1}{\rm Mpc}$ box size and a resolution $N_{\rm p}=192^3$ particles per type;
bottom panels display slices of the internal energy of the gas, for the same redshift interval. 
Although the two cosmologies are rather different, our remapping (\ref{eq3}-\ref{eq5}) produces
almost identical nonlinear total matter and flux power spectra, and therefore 
a similar LSS morphology with no  perceptible visual differences in the cosmic web structure.}
\label{fig:nonstandard_sims} 
\end{figure*}

%---------------------------------------------------------------------------------------------------------------

\subsection{Approximation for Dark Radiation Models}

To account for non-standard dark radiation scenarios in $\cal{L}^{\rm Ly\alpha}$, we
extended the parameter space $\boldsymbol{\Theta}$ to include models with sterile neutrinos or more generic relic radiation, where $N_{\rm eff}$ is 
different from the canonical reference value corresponding to three thermalized active neutrinos (i.e., $N_{\rm eff}=3.046$). The Taylor expansion
of the one-dimensional Ly$\alpha$ flux power spectrum will then  include further terms, due to the presence of  a non-standard $N_{\rm eff}$ value, but the logic leading to the 
construction of $\mathcal{L}^{\rm Ly\alpha}$ remains essentially the same. Hence, in principle we just require additional cosmological hydrodynamical simulations to map out 
the extended parameter space and evaluate extra cross-derivative terms in the Taylor expansion.
However, this computationally expensive procedure can be avoided  
 with the following strategy. 
Consider two models $\cal{M}$ and $\tilde{\cal{M}}$ defined by $N$ cosmological parameters 
$\boldsymbol{\alpha}$ and $\boldsymbol{\tilde{\alpha}}$, which also include massive neutrinos. 
Model $\cal{M}$ is the reference model with the standard value of $N_{\rm eff}=3.046$, while  model $\tilde{\cal{M}}$   has $ \tilde{N}_{\rm eff} = N_{\rm eff}+ \Delta N_{\rm eff}$, with $\Delta N_{\rm eff} \ne 0$.
We restrict our analysis to the case of 
three species of degenerate massive neutrinos and  
assume individual neutrino masses $m_{\rm \nu, i} < 0.6$ eV, so that they are fully relativistic
at the redshift of equality $z_{\rm eq}$. The basic idea is to map the model $\cal{M}$ 
into a different model  $\tilde{\cal{M}}$ with $N_{\rm eff} \ne 3.046$, which produces the same (or almost the same) total matter linear power spectrum as $\cal{M}$. 
If the two models are characterized by the same linear matter power spectrum, they will also have nearly identical
nonlinear matter and flux power spectra. Hence, one can simply rely on linear theory and 
on simulations with standard $N_{\rm eff}$ to specify more exotic dark radiation scenarios.
In practice, there should also be a small effect due to the fact that the expansion rate changes with $N_{\rm eff}$, but
this effect is ignored here since we 
 neglect radiation density in our simulations.
It is easy to prove that the previous condition is realized if $\cal{M}$ and $\tilde{\cal{M}}$ have the same 
values of $z_{\rm eq}$, $\Omega_{\rm m}$, $\omega_{\rm b}/\omega_{\rm c}$ and $f_{\rm \nu}$, 
with $\Omega_{\rm m}$ the matter density, 
$\omega = \Omega h^2$,  and  $f_{\rm \nu} = \omega_{\rm \nu} / \omega_{\rm m}$ -- where the labels $m, b, c, \nu$ stand for
total matter, baryons, cold dark matter, and neutrinos -- respectively. This is true up  to small differences in the scale of BAO peaks, 
but the fact that the location of BAOs slightly differs in the two cases is unimportant for the Ly$\alpha$ likelihood. 
In particular, the condition on $f_{\nu}$ guarantees that both the small-scale suppression in the matter power spectrum  and
the small-scale linear growth factor are identical in $\cal{M}$ and $\tilde{\cal{M}}$. 
Based on these requirements, the following two models will have
nearly the same total linear  matter power spectrum:
\begin{eqnarray} 
{\cal{M}} &=& \{ \omega_{\rm b}, \omega_{\rm c}, H_0, N_{\rm eff}, \omega_{\rm \nu} \} \label{eq3} \\
 \tilde{\cal{M}} &=& \{ \tilde{\omega}_{\rm b}, \tilde{\omega}_{\rm c}, \tilde{H}_0, \tilde{N}_{\rm eff}, \tilde{\omega}_{\rm \nu} \}  \nonumber \\
  &=& \{ \eta^2 \omega_{\rm b}, \eta^2 \omega_{\rm c}, \eta H_0, N_{\rm eff}+\Delta N_{\rm eff}, \eta^2 \omega_{\rm \nu} \}
\label{eq4}
\end{eqnarray}
with  
\begin{equation}
\eta^2 = [1 + 0.2271 (N_{\rm eff}+\Delta N_{\rm eff} ) ] / [1+0.2271 N_{\rm eff}]
 \label{eq5}
\end{equation}
and $\tilde{M}_{\rm \nu} =  \tilde{M}_{\rm \nu}^{\rm a}  + \tilde{M}_{\rm \nu}^{\rm s} = \eta^2 M_{\rm \nu}$ -- where in the last passage we distinguish between
the active and sterile contributions to the total mass (if the sterile neutrino has non-zero mass), and $M_{\rm \nu} = \sum m_{\rm \nu}$.
To this end, in terms of structure formation there is no actual difference if the total mass is given by a combination of
active or sterile neutrinos, or just by active neutrinos for example -- since what is really relevant is eventually the total neutrino number density. 
However, our formalism is more general and can also account for the mass fraction of a sterile neutrino if the latter one is assumed to be massive. 

Figure \ref{fig1} shows that   the previous approximation is accurate within $1\%$
in the regime of interest (i.e., BOSS Ly$\alpha$ forest region, shaded cyan area in the left panel),  which is comparable with our expected uncertainties from hydrodynamical simulations (see the next section). 
In essence, the figure illustrates a 1\% difference between the remapping and the exact formulation when one considers linear evolution, which we take as our systematics.
Specifically, the  
left panel shows linear power spectra computed with CAMB \cite{CAMB2000} for different dark radiation models $\tilde{\cal{M}}$ having $\Delta N_{\rm eff}=1$ at $z=3$ (chosen as 
an indicative central redshift value for our simulations), 
normalized by the baseline model $\cal{M}$ which has $N_{\rm eff}=3.046$ and assumes three active neutrinos of degenerate mass  -- when $M_{\rm \nu}=0.3$ eV.
In particular, model A1 -- characterized by a massless sterile neutrino thermalized with three active neutrinos of degenerate mass -- is the main focus of this study, while in the other models
the sterile neutrino is massive, thermalized, and shares the same mass as the three active species (B1), or has a different mass (C1); in the latter case, the mass fraction of the sterile neutrino is $(1-\eta^{-2})$ of the
total neutrino mass of the baseline model. Note that the small tilt between models is mainly due to whether or not the sterile neutrino is assumed to be massive,
and therefore differences between A1 and either B1 or C1 are more pronounced (rather than between B1 and C1).
The right panel shows the CMB power spectra for the same models, which are significantly different -- unlike the linear matter power spectra.
At higher redshift and up to the time of radiation-to-matter equality, the difference between the various linear power spectra is as small as at $z=0$.
Our strategy is to use this analytic approximation only in the Ly$\alpha$ likelihood; for the CMB and BAO scale likelihoods, we always assume the full exact models.
  
%%%%%%%%%%%%%%%%%%%%%%%%%%%%%%%%%%%%%%%%%%%%%%%%%%%%%%%%%%%%%%%%%%%%%%%%%%%%%%%%%%%%%%%%%%%%%%%%%%%%%%%%%%%%%%%%%%%%%%%%%%%%%%%%%%%%%%%%%%%%%%%%%%%%%%%%%%%%%%%%%

% RESULTS

\section{Results} \label{sec:results}

%---------------------------------------------------------------------------------------------------------------

\begin{figure}
\includegraphics[scale=0.375]{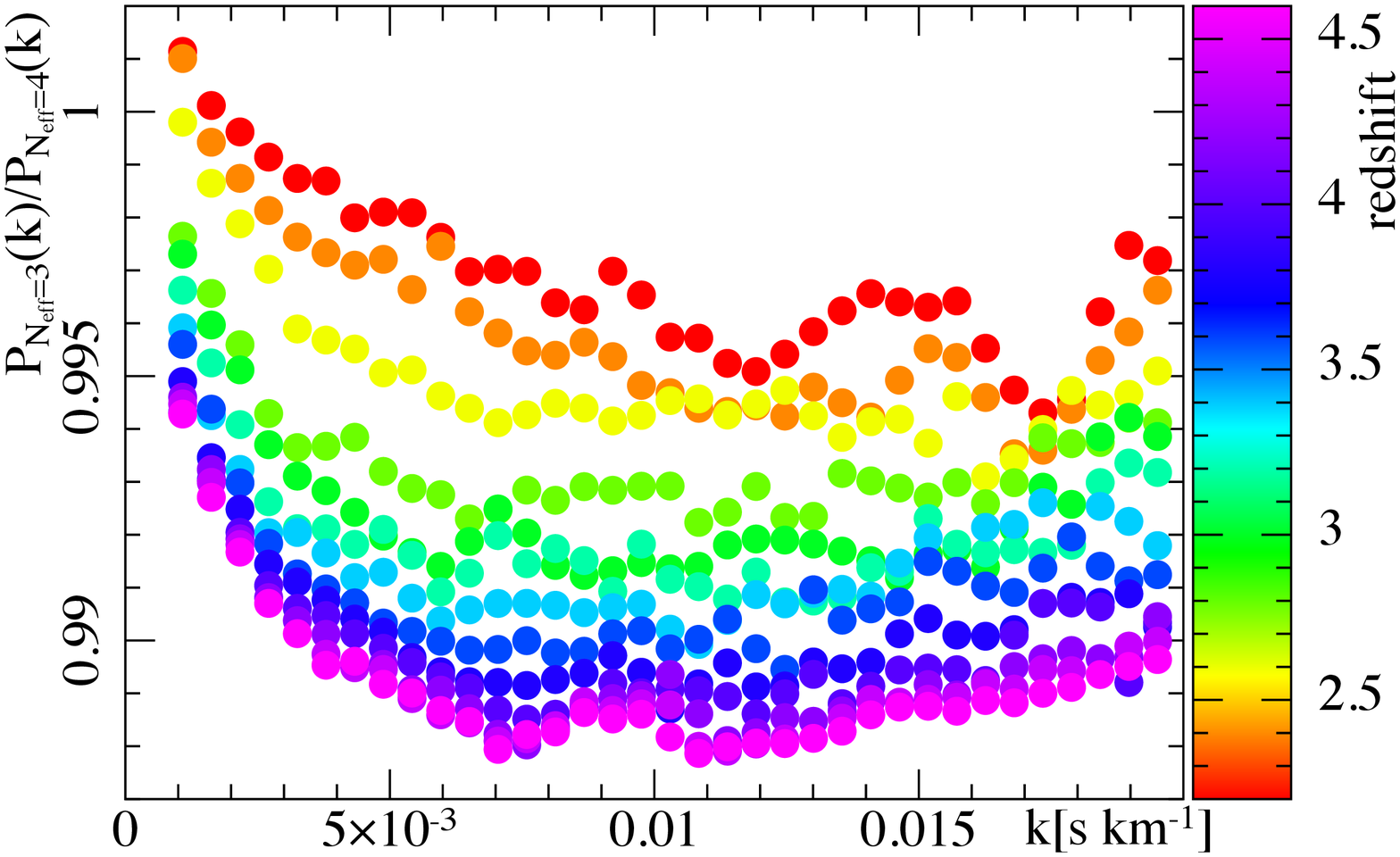}
\caption{Ratios of synthetic
one-dimensional Ly$\alpha$ flux power spectra extracted from a
baseline model $\cal{M}$ having three degenerate massive neutrinos and no extra relativistic degrees of freedom  ($N_{\rm eff}=3$, $M_{\rm \nu}=0.35$ eV),  
and from a non-standard dark radiation model $\tilde{\cal{M}}$ characterized by a massless sterile neutrino and three
active neutrinos of degenerate mass ($\tilde{N}_{\rm eff}=4$, $\tilde{M}_{\rm \nu}=0.4$ eV). 
The cosmological parameters of $\cal{M}$ and $\tilde{\cal{M}}$ 
 are fixed according to (\ref{eq3}) and (\ref{eq4}).
At any given redshift, indicated by different colors in the figure, deviations in the corresponding power spectra 
are all within $1\%$ (comparable to 
those obtained from linear theory), validating our analytic remapping also in the nonlinear regime.}
\label{fig2} 
\end{figure}

%---------------------------------------------------------------------------------------------------------------

\begin{figure}
\includegraphics[scale=0.47]{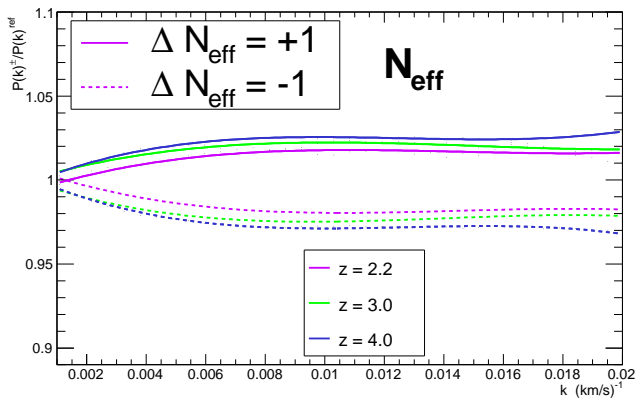}
\caption{Sensitivity of the Ly$\alpha$ flux power spectrum to a variation in $N_{\rm eff}$.  
Assuming a central reference value $N_{\rm eff}=3$, the diagram illustrates that a change
$\Delta N_{\rm eff} = \pm 1$  in    $N_{\rm eff}$  (solid or dotted lines in the figure, respectively)
produces a global change in the  Ly$\alpha$   flux up to 3\%  at the representative redshifts considered,
more significant than the uncertainty
associated with our simulations  or with our dark radiation approximation (Eqns. \ref{eq3}-\ref{eq5}).} 
\label{fig3} 
\end{figure}

%---------------------------------------------------------------------------------------------------------------

\begin{figure}
\includegraphics[scale=0.44]{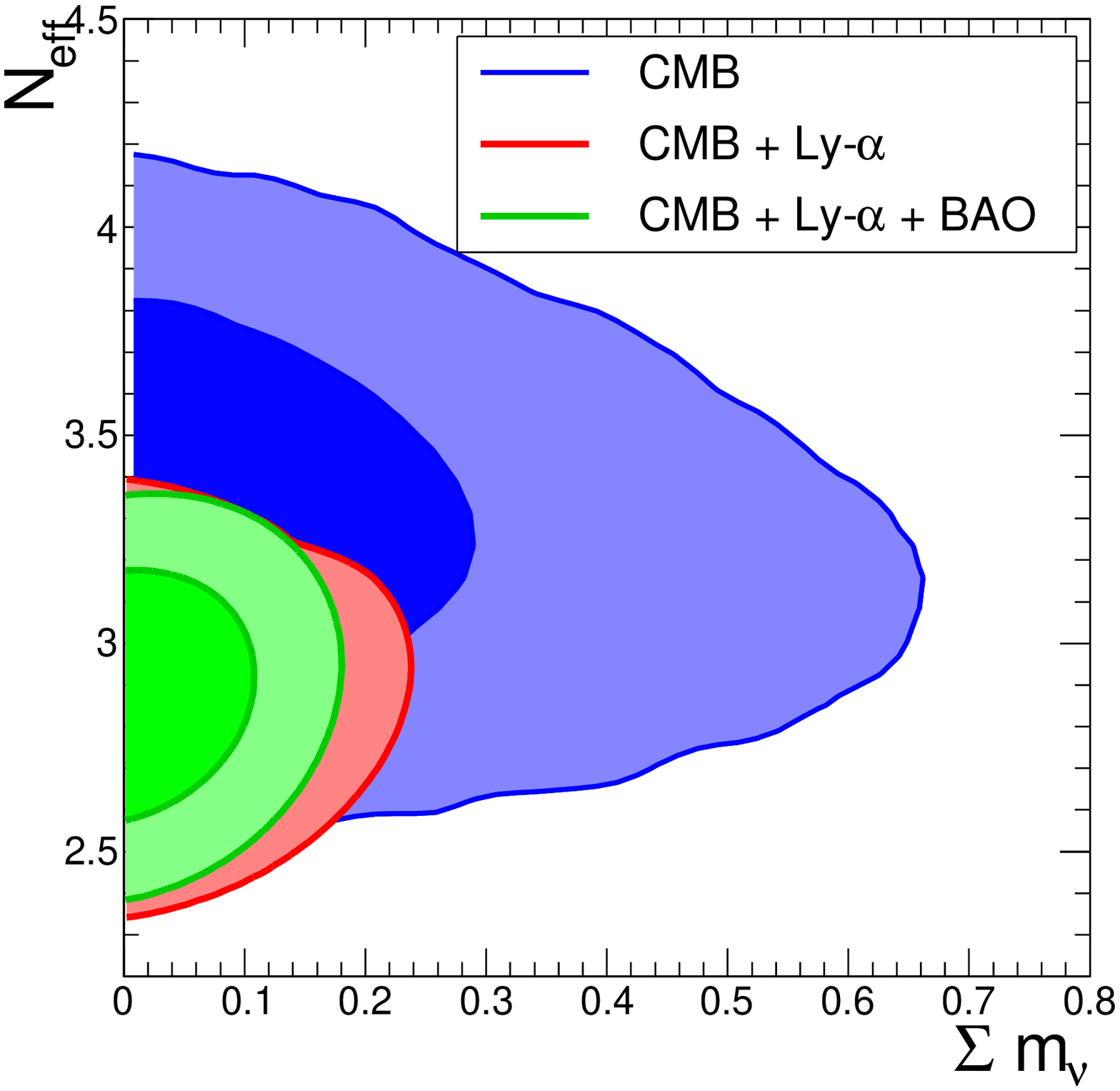}
\caption{Joint constraints on the number of effective neutrino species $N_{\rm eff}$
and the total neutrino mass $\sum m_{\rm \nu}$,  obtained from different
cosmological probes. Red contours refer to the combination of CMB+Ly$\alpha$
data, while green contours include additional information from BAOs;
in the first case we obtain
$N_{\rm eff}= 2.91^{+ 0.21}_{- 0.22}$  
and $\sum m_{\rm \nu} <0.15$ eV, while in the second  
$N_{\rm eff}= 2.88 \pm 0.20$ 
and $\sum m_{\rm \nu} <0.14$ eV -- all at 95\% CL.
Our results exclude the possibility of a sterile neutrino -- thermalized with active neutrinos  -- at a significance of over  5 $\sigma$, and
provide strong evidence for the CNB from $N_{\rm eff}\sim 3$ -- as $N_{\rm eff} = 0$ is rejected at more than $14~\sigma$.}
\label{fig4}
\end{figure}

%---------------------------------------------------------------------------------------------------------------

The accuracy of our analytic approximation (Eqs. \ref{eq3}-\ref{eq5}) to include non-standard dark radiation models in the Ly$\alpha$ likelihood
 has also been tested in the nonlinear regime, by 
performing cosmological hydrodynamical simulations with non-standard $N_{\rm eff}$ values and verifying the robustness of our fitting procedure -- along with the correct recovery of 
the nonlinear matter and Ly$\alpha$ flux power spectra.
For example, we run a simulation based on a model $\tilde{\cal{M}}$ with 
$\tilde{N}_{\rm eff}=4$ and $\tilde{M}_{\rm \nu}=0.4$ eV, where an additional massless sterile neutrino
is assumed to be in thermal equilibrium with three degenerate active massive neutrinos; we also run the
corresponding baseline model $\cal{M}$  having $N_{\rm eff}=3$ and $M_{\rm \nu}= \tilde{M}_{\rm \nu}/\eta^2=0.35$ eV -- where the cosmological parameters are determined according to
(\ref{eq3}) and (\ref{eq4}).  
Figure \ref{fig:nonstandard_sims} shows selected snapshots at $z=3$ from those simulations:
left panels refer to the  baseline model $\cal{M}$, while right panels are for the
non-standard model $\tilde{\cal{M}}$. In the top panels, we display  projections of the gas density
along the $x$ and $y$ directions (and across $z$) for a
$25~h^{-1} {\rm Mpc}$ box size and a resolution $N_{\rm p}=192^3$ particles per type;
in the bottom ones, we show slices of the internal energy of the gas for the same redshift interval. 
The axis scales are in $h^{-1 }{\rm Mpc}$, and the various plots are 
smoothed with a cubic spline kernel. The main point of the plot  is to provide a visual proof that
essentially models with rather different cosmologies, mapped through our approximation, 
eventually produce almost identical nonlinear total matter and flux power spectra, and therefore present
nearly the same LSS morphology (differences are not visually perceptible).   

Figure \ref{fig2} is a further confirmation that the discrepancy between the exact formulation and our
approximated  remapping procedure  is not worst than 1\% even in the nonlinear regime, thus at the same level of 
agreement as the aforementioned assumed systematic uncertainty when we consider the linear regime. 
The figure shows  the ratios of  synthetic
Ly$\alpha$ forest flux power spectra extracted at different redshifts from those two models (note that this is not the expected signal): even in the nonlinear regime, 
we find that deviations in the power spectra of  $\tilde{\cal{M}}$ and $\cal{M}$ are within $1\%$ 
for all the $z$-intervals of interest.

It is also useful to quantify the total variation in the Ly$\alpha$ forest flux power spectrum due to a change in $N_{\rm eff}$,
in order to assess the magnitude of the 
systematic error related  to our remapping procedure -- along with the
uncertainty associated with our simulations -- relative to the magnitude of the change in $N_{\rm eff}$ we are seeking.
For some representative redshift values (i.e., z=2.2, 3.0, 4.0), Figure \ref{fig3} shows that a variation $\Delta N_{\rm eff} = \pm 1$  in    $N_{\rm eff}$ from its central value (assumed to be $N_{\rm eff}=3$), 
indicated respectively with solid or dotted lines in the figure,
 causes a global change in the  Ly$\alpha$   flux   from 2.2\% up to 3\%  at the various redshifts considered. 
Therefore, the effect we are looking for is more significant than our  uncertainty
associated with simulations,  or with our dark radiation approximation (both within 1\% level). 
In \cite{PDB2015}, the response of the Ly$\alpha$ flux  to
isolated variations in other individual parameters, such as the total neutrino mass, $\sigma_8$, the spectral index $n_{\rm s}$, $H_0$ or $\Omega_{\rm m}$,
has been quantified (see their Figure 12 and their Section 5.1), and coherent percent-level changes in
the power spectrum across multiple $k$-bins
 and redshift slices have been proven to be detectable with very high statistical significance. 
 
Having fully validated our analytic approximation and quantified the sensitivity of 
the Ly$\alpha$ flux power spectrum to a variation in $N_{\rm eff}$, we implemented the extension to dark radiation models in 
the procedure applied in \cite{PDB2015}, and then interpreted the  global likelihood $\cal{L}$
in the context of the frequentist approach \cite{NEY1937} -- along the lines explained in Section \ref{sec:frequentist}.
Figure \ref{fig4} summarizes the main results of our fitting procedure for the values of 
$N_{\rm eff}$ and $\sum m_{\rm \nu}$, derived by combining
CMB (Planck+ACT+SPT+WMAP polarization; blue contours) with Ly$\alpha$ forest data (red contours),
or by further adding BAO information (green contours). 
Specifically,  
we obtain $N_{\rm eff}= 2.91^{+ 0.21}_{- 0.22}$ (95\% CL) 
and $\sum m_{\rm \nu} <0.15$ eV (95\% CL) in the first case,
and $N_{\rm eff}= 2.88 \pm 0.20$ (95\% CL) 
and $\sum m_{\rm \nu} <0.14$ eV (95\% CL) in the second. 
Table \ref{tab1} reports the final results of the 
fits for all the
main cosmological parameters ($\boldsymbol{\alpha}$), in addition to 
$N_{\rm eff}$ and $\sum m_{\rm \nu}$, for
the two combinations of datasets 
considered (i.e., CMB+Ly$\alpha$ or CMB+Ly$\alpha$+BAO). 
In particular, our tight constraints on $N_{\rm eff}$
exclude the possibility of a sterile neutrino thermalized with active neutrinos -- or more generally 
of any decoupled relativistic relic with $\Delta N_{\rm eff} \simeq 1$
-- at significance of over  5 $\sigma$, the strongest bound to date, and  are fully consistent with the latest constraints 
recently reported by Planck (2015) \cite{Planck_2015_parameters}.
We discuss the major implications of these results in cosmology and particle physics next. 

%---------------------------------------------------------------------------------------------------------------

\begin{table}[t]
\begin{center}
\begin{tabular}{lcc}
\hline
Parameter &   CMB+Ly$\alpha$ &  CMB+Ly$\alpha$+BAO  \\
\hline 
$n_{\rm s}$   &  $0.950_{-0.008}^{+0.007}  $ &  $0.949 \pm 0.007   $   \\
$H_0$~{\scriptsize[km/s/Mpc]}    & $67.0 \pm 1.3$  & $66.8 \pm 1.3$  \\
$\sum \! m_\nu$~{\scriptsize[\rm eV]}   & $<0.15$ {\scriptsize (95\%)}  & $<0.14$ {\scriptsize (95\%)} \\
$\sigma_8$  & $0.831_{-0.015}^{+0.013} $   & $0.834_{-0.020}^{+0.015} $ \\
$\Omega_{\rm m}$   &  $0.308\pm 0.015$  &  $0.311\pm 0.009$ \\
$N_{\rm eff}$ & $2.91_{-0.22}^{+0.21}$ &   $2.88 \pm 0.20 $  \\
\hline
\end{tabular}
\caption{Values of the main cosmological parameters obtained from a frequentist analysis of the likelihood $\cal{L}$, as explained in the main text, for the
two combinations of datasets considered in this work -- CMB+Ly$\alpha$ or CMB+Ly$\alpha$+BAO.}
\label{tab1}
\end{center}
\end{table}

%%%%%%%%%%%%%%%%%%%%%%%%%%%%%%%%%%%%%%%%%%%%%%%%%%%%%%%%%%%%%%%%%%%%%%%%%%%%%%%%%%%%%%%%%%%%%%%%%%%%%%%%%%%%%%%%%%%%%%%%%%%%%%%%%%%%%%%%%%%%%%%%%%%%%%%%%%%%%%%%%

% DISCUSSION

\section{Discussion} \label{sec:conclusions}

%---------------------------------------------------------------------------------------------------------------

Simultaneous constraints on   $N_{\rm eff}$ and $\sum m_{\rm \nu}$ are
interesting, since extra relics could coexist with massive neutrinos or could themselves have a mass in the eV range.
From CMB measurements alone, these two parameters
 do not show significant correlations because their physical effects can be
resolved individually, while $N_{\rm eff}$ and $\sum m_{\rm \nu}$ may be  partially degenerate when considering LSS tracers
-- actually, in the range of validity of the analytic approximation that we use for Ly$\alpha$ data, these two parameters are totally degenerate, but
outside of that regime the two quantities may not be fully degenerate because of different effects on the expansion rate and growth factor. 
As reported in \cite{PDB2015}, the main correlation observed is between $\sum m_{\rm \nu}$ and $\sigma_8$, where
the correlation coefficient reaches $\sim 70 \%$; also, for a fixed $\sigma_8$ the change in the Ly$\alpha$ flux power spectrum is
a nearly constant $1\%$ increase at $z=4.0$, while at $z=2.2$ the change declines to nearly zero (see again their Figure 12 and their Section 5.1).  
Similarly, in the region of interest here $N_{\rm eff}$ is also essentially degenerate with $\sigma_8$ (compare our Figure 4 with Figure 12 in \cite{PDB2015} for $\sigma_8$). 
However, interestingly enough, this degeneracy can be broken at larger scales, for instance by considering voids as LSS tracers and their properties in relation to massive neutrinos. 
Indeed,  the limiting factor in cosmological constraints is the ability to break degeneracies among the effects of different parameters.
 However,  the most constraining  power
comes from the combination of CMB and LSS,
because distinct cosmological probes have  different and independent systematic errors, and contrasting directions
of degeneracy in parameter space. This is particularly true for the Ly$\alpha$ forest, which reduces the uncertainties on cosmological parameters quite significantly 
when combined with CMB measurements.
With respect to the total neutrino mass, the 
 ability to place a strong upper limit  ultimately derives from the fact that 
 the distinctive scale- and redshift-dependence suppression of power  in the matter and Ly$\alpha$ flux power spectrum caused by neutrinos 
 cannot be mimicked by a combination of other parameters, and is not fully degenerate.
 In the case of $N_{\rm eff}$, most of the information  comes from precise measurements of the photon diffusion scale relative to the sound horizon scale (hence from the CMB), but
the combination of other parameters in the Ly$\alpha$ likelihood and very different directions
of degeneracy in parameter space contribute to tighter limits. 
For example, we tested this by completely
removing the dependence on $N_{\rm eff}$ in $\mathcal{L}^{Ly\alpha}$, and 
found that our final limits  on $N_{\rm eff}$ varied only marginally -- confirming that 
most of the constraining power on the number of effective neutrino species indeed resides in the CMB, although 
some additional -- albeit small -- information is also contained in the Ly$\alpha$ forest.
Therefore, we would expect that the combination of CMB+Ly$\alpha$ will always perform better than the CMB alone, 
and if combined with the new Planck (2015) data \cite{Planck_2015_overview} the results presented here will be even tighter. 
In essence, the key is the synergy of the CMB with a high-redshift tracer having different systematics and probing
 different directions in parameter space. We also note that  
there is no significant correlation between  $N_{\rm eff}$ and $\sum m_{\rm \nu}$ in the CMB+Ly$\alpha$ contours 
(essentially because the CMB is driving the constraints particularly on $N_{\rm eff}$, and as previously mentioned from CMB
measurements alone these two parameters do not show significant correlations), and therefore
our upper limits on the total neutrino mass obtained from a joint analysis 
are consistent with \cite{PDB2015}.

There has long been a worry about being able to trust Ly$\alpha$ forest data and extract the power spectrum at small scales.
However, in the recent few years the statistical power of the forest has increased dramatically thanks to new exquisite data from the SDSS survey, and in particular 
to BOSS \cite{SDSS2000,BOSS2013}. The situation will certainly improve with eBOSS and DESI  \cite{DESI2013,ABA2015}. Along with better and higher-quality data, our understanding of the various systematics affecting the Ly$\alpha$ forest
has also improved significantly over the past few years. 
To this end, \cite{PDB2013} has conducted a careful analysis of Ly$\alpha$ forest data from BOSS, and accounted for a
long list of systematic effects. From the numerical side, \cite{ROS2014}
 performed a detailed analysis of the modeling of the
small-scale Ly$\alpha$ flux spectrum in presence of massive neutrinos, quantifying (from a pure theoretical ground) the impact of systematics at small-scales. 
In addition,  \cite{PDB2015} presented an accurate analysis on small-scale systematics.    
In general, a rather conservative assessment of the neutrino mass limits has been performed, particularly regarding the splicing technique.
 The impact of the major known systematics have been quantified and taken into account in our technique with a series of nuisance
 parameters. 
 In particular, UV fluctuations, AGN and SNe feedback, high-density absorbers, point spread function (PSF) of BOSS spectra
and  splicing uncertainties have been  translated in terms of uncertainties in our quoted limits on $N_{\rm eff}$ and $\sum m_{\rm \nu}$.  
However, even after accounting for all the possible known systematics, 
a small tension between Ly$\alpha$ forest and CMB data remains, namely the power
spectrum amplitude obtained from the former probe is somewhat larger than that preferred by the latter one, and this fact may
be driving our tighter limits; further investigation along these lines is ongoing work.  

Joint constraints on the number of effective  neutrino species and the total neutrino mass 
are also in general model-dependent.  In this study, to derive our limits on 
 $N_{\rm eff}$ and $\sum m_{\rm \nu}$ we  assumed that the three active neutrinos share a mass of $\sum m_{\rm \nu}/3$,
where $m_{\rm \nu, i} < 0.6$ eV, 
and may coexist with massless extra species contributing to $N_{\rm eff}$ as $\Delta N_{\rm eff}$. 
Based on these assumptions, the main conclusions of our analysis are as follows:
(1)  the possibility of a sterile neutrino thermalized with active neutrinos -- or more generally 
of any decoupled relativistic relic with $\Delta N_{\rm eff} \simeq 1$
-- is ruled out at a significance of over  5 $\sigma$, the strongest bound to date; 
(2) as in \cite{PDB2015}, we obtain a tight and competitive upper bound on the total neutrino mass, which eventually will be helpful in solving the neutrino hierarchy problem; 
(3) by rejecting $N_{\rm eff} = 0$ at more than $14~\sigma$, our constraints provide the strongest evidence for the CNB from $N_{\rm eff} \sim 3$.
These results have several important implications in particle physics and cosmology. In particular, 
 the effective number of neutrino-like relativistic degrees of freedom is found compatible with the
canonical value of 3.046 at high-confidence, suggesting that 
the minimal $\Lambda$CDM model -- along with its thermal history -- is strongly favored 
over extensions with non-standard  neutrino properties or with extra-light degrees of freedom,
and the measured energy density is composed of standard model neutrinos.  Hence, no new neutrino physics nor new particles are required, and
the theoretical assumptions going into the standard cosmology theory are correct. In addition, along with \cite{PDB2015}, our stringent upper 
bounds on $\sum m_{\rm \nu}$  
suggest interesting complementarity with future particle physics 
direct measurements of the effective electron neutrino mass  \cite{CAP2014}.
Finally, our conclusions on the CNB nicely complement recent results from Planck (2015), which has 
detected the free-streaming   nature of the species responsible for $N_{\rm eff}\sim3$
with high significance \cite{Planck_2015_parameters,Planck_2015_inflation}. 
We expect that our joint constraints on  $N_{\rm eff}$ and $\sum m_{\rm \nu}$   will be improved by a factor of 2 by including eBOSS measurements, while DESI should improve
these constraints even further \cite{DESI2013,FR2014,ABA2015} -- and likely shed a novel light into the hierarchy nature of the masses of active neutrinos, along with the 
reconstruction of the individual mass of each neutrino mass eigenstate. 

%%%%%%%%%%%%%%%%%%%%%%%%%%%%%%%%%%%%%%%%%%%%%%%%%%%%%%%%%%%%%%%%%%%%%%%%%%%%%%%%%%%%%%%%%%%%%%%%%%%%%%%%%%%%%%%%%%%%%%%%%%%%%%%%%%%%%%%%%%%%%%%%%%%%%%%%%%%%%%%%%

% ACKNOWLEDGEMENT

\acknowledgements 
This work is supported by the National Research Foundation of Korea (NRF) 
through NRF-SGER 2014055950 funded by the Korea government (MEST), and by
the faculty research fund of Sejong University in 2014. Some numerical simulations developed for this study were
 performed using the Korea Institute of Science and Technology Information (KISTI) supercomputer (Tachyon-I) under allocation  KSC-2014-C1-045. 
 We also acknowledge PRACE for awarding us access to resource Curie-thin nodes based in France at TGCC.
N.P.-D. and Ch.Y. acknowledge support from grant ANR-11-JS04-011-01 of Agence Nationale de la Recherche.  

%%%%%%%%%%%%%%%%%%%%%%%%%%%%%%%%%%%%%%%%%%%%%%%%%%%%%%%%%%%%%%%%%%%%%%%%%%%%%%%%%%%%%%%%%%%%%%%%%%%%%%%%%%%%%%%%%%%%%%%%%%%%%%%%%%%%%%%%%%%%%%%%%%%%%%%%%%%%%%%%%

%  BIBLIOGRAPHY

%%%%%%%%%%%%%%%%%%%%%%%%%%%%%%%%%%%%%%%%%%%%%%%%%%%%%%%%%%%%%%%%%%%%%%%%%%%%%%%%%%%%%%%%%%%%%%%%%%%%%%%%%%%%%%%%%%%%%%%%%%%%%%%%%%%%%%%%%%%%%%%%%%%%%%%%%%%%%%%%%
%%%%%%%%%%%%%%%%%%%%%%%%%%%%%%%%%%%%%%%%%%%%%%%%%%%%%%%%%%%%%%%%%%%%%%%%%%%%%%%%%%%%%%%%%%%%%%%%%%%%%%%%%%%%%%%%%%%%%%%%%%%%%%%%%%%%%%%%%%%%%%%%%%%%%%%%%%%%%%%%%

\end{document}